\documentclass[aps,prc,twocolumn,floatfix,nofootinbib,showpacs,superscriptaddress]{revtex4-1}

\usepackage{amsmath,amsfonts,amssymb,bm}
\usepackage{graphicx}
\usepackage{color}
\usepackage{pzccal}
\usepackage{soul}
\definecolor{purple}{rgb}{0.5,0,0.5}
\definecolor{blue}{rgb}{0.0,0,0.9}

\begin{document}

\title{Dressed-quarks and the nucleon's axial charge}

\author{Lei Chang}
\affiliation{Institut f\"ur Kernphysik, Forschungszentrum J\"ulich, D-52425 J\"ulich, Germany}

\author{Craig D.~Roberts}
\affiliation{Institut f\"ur Kernphysik, Forschungszentrum J\"ulich, D-52425 J\"ulich, Germany}
\affiliation{Physics Division, Argonne National Laboratory, Argonne, Illinois 60439, USA}
\affiliation{Department of Physics, Illinois Institute of Technology, Chicago, Illinois 60616-3793, USA}

\author{Sebastian M.~Schmidt}
\affiliation{Institute for Advanced Simulation, Forschungszentrum J\"ulich and JARA, D-52425 J\"ulich, Germany}


\date{11 September 2012~}

\begin{abstract}
The nucleon's axial charge, $g_A$, expresses features that are both fundamental to the strong interaction and crucial to its connection with weak interaction physics.
We show that dynamical chiral symmetry breaking (DCSB) suppresses the axial-charge of a dressed-quark, $g_A^q$, at infrared momenta.  Since this effect disappears as chiral symmetry is restored, one may argue that $g_A$ vanishes with the restoration of chiral symmetry because no nucleon bound-state survives the associated transition.
The suppression of $g_A^q$ is shown to be part of an explanation for a 25\% reduction of $g_A$ from its nonrelativistic quark-model value.  Critical too, however, is the presence of dressed-quark angular momentum within the nucleon.
The value of $g_A^q$ depends on the kernels of the gap and Bethe-Salpeter equations.  We find that incorporation of essentially nonperturbative effects associated with DCSB into these kernels inflates the value relative to that obtained at leading-order in a widely used truncation of QCD's Dyson-Schwinger equations.
Such corrections also affect the nucleon's axial radius.  In both cases, however, agreement with experiment will require similar improvements to the Faddeev kernel and associated interaction current.
\end{abstract}


\maketitle

\section{Introduction}
The prototypical weak interaction is nuclear $\beta^-$-decay, which explains the instability of neutron-rich nuclei and proceeds via the transition
\begin{equation}
n\to p + e^- + \bar \nu_e\,.
\label{betadecay}
\end{equation}
The first attempt at its explanation \cite{Fermi:1934sk,Fermi:1934hr} was based on a contact current-current interaction, modulated by a constant \cite{Nakamura:2010zzi}: $G_F = 1.166 \times 10^{-5}\,{\rm GeV}^{-2}$.  Electroweak gauge theory replaces the contact interaction by exchange of a heavy gauge boson and
produces the tree-level expression $(G_F/\surd 2) = g^2 / (8 M_W^2)$, where $M_W \sim 80\,$GeV is the gauge-boson's mass and $g$ is a universal dimensionless coupling; namely, it is the same for all interactions between gauge-bosons, leptons and current-quarks.

Neutron $\beta$-decay and kindred processes play a crucial role in many domains, e.g.: Big-Bang nucleosynthesis, constraining the abundance of deuterium; supernovae explosions, producing a vast amount of energy through neutrino production; testing the Standard Model, placing constraints on extensions via low-energy experiments; and in practical applications, such as $\!^{14}$C-dating and positron emission tomography.  Notwithstanding its widespread importance, a connection between the coupling, $g$, that describes weak processes involving current-quarks and that between weak bosons and the dressed-quarks produced by nonperturbative interactions in QCD, the strongly interacting part of the Standard Model, has not been elucidated.  Attempts to do so are described in Refs.\,\cite{Edwards:2005ym,Renner:2006sh,Roberts:2007jh,Yamazaki:2008py,Eichmann:2011pv, Renner:2010ks,Wittig:2012ha,Hall:2012qn}.

The transition in Eq.\,\eqref{betadecay} may be studied via the quark-based axial-vector matrix element
\begin{equation}
\Lambda_{5\mu}^{pn}(p_f,p_i) = \langle p(p_f,\lambda_f) |\, \bar u \gamma_5\gamma_\mu d\, | n(p_i,\lambda_i)\rangle
\end{equation}
where $p_{i,f}$, $\lambda_{i,f}$ are, respectively, initial/final momenta and helicities associated with the initial-state neutron and final-state proton.  If one assumes isospin symmetry, then Poincar\'e covariance entails that this matrix element is completely described by two form factors \cite{Schindler:2006jq}:
\begin{eqnarray}
\nonumber
\Lambda_{5\mu}^{pn}(p_f,p_i) &=& \bar u_p(p_f,\lambda_f)
\bigg[ \gamma_5\gamma_\mu \, G_A(q^2) \\
&& \quad  +\,  i\gamma_5 \frac{1}{2M_N}\, q_\mu \,G_P(q^2) \bigg] u_n(p_i,\lambda_i)\,,
\end{eqnarray}
where $q=p_f-p_i$, $G_A(q^2)$ is the nucleon's axial-vector form factor, $G_P(q^2)$ is its induced pseudoscalar form factor and $M_N$ is the average nucleon mass.\footnote{We use a Euclidean metric:  $\{\gamma_\mu,\gamma_\nu\} = 2\delta_{\mu\nu}$; $\gamma_\mu^\dagger = \gamma_\mu$; $\gamma_5= \gamma_4\gamma_1\gamma_2\gamma_3$, tr$[\gamma_5\gamma_\mu\gamma_\nu\gamma_\rho\gamma_\sigma]=-4 \epsilon_{\mu\nu\rho\sigma}$; $\sigma_{\mu\nu}=(i/2)[\gamma_\mu,\gamma_\nu]$; $a \cdot b = \sum_{i=1}^4 a_i b_i$; and $P_\mu$ timelike $\Rightarrow$ $P^2<0$.}  The quantity of primary interest herein is the nucleon's nonsinglet axial-charge
\begin{equation}
g_A:=G_A(q^2=0)\,.
\end{equation}

The axial charge also has a relation to nucleon spin physics.  Given our assumption of isospin symmetry, then
\begin{eqnarray}
\nonumber
\lefteqn{
\langle p(p_f,\lambda_f) |\, \bar u \gamma_5\gamma_\mu d\, | n(p_i,\lambda_i)\rangle}\\
&=&
2\, \langle p(p_f,\lambda_f) |\,
\bar u \gamma_5\gamma_\mu u
-\bar d \gamma_5\gamma_\mu d \, | p(p_i,\lambda_i)\rangle \,. \label{isosymmetry}
%
\end{eqnarray}
In the forward scattering limit; i.e., $p_f=p_i=p$, with $\lambda_f=\lambda_i= \lambda$,  $\lambda\cdot p = + \mbox{\small $\frac{1}{2}$}$, then in the infinite-momentum frame
\begin{eqnarray}
2 M_N \,  \lambda_\mu \, \langle q_\uparrow \rangle & = &
\langle p(p,\lambda) |\, \bar q \gamma_5\gamma_\mu q \, | p(p,\lambda)\rangle \,,\\
\langle q_\uparrow \rangle &=& \int_0^1 \!dx\, \left[ \Delta q(x) + \Delta \bar q(x) \right]\,,
\end{eqnarray}
where $\Delta q(x) = q_\uparrow(x) - q_\downarrow(x)$ is the light-front helicity distribution for a quark $q$ carrying a fraction $x$ of the proton's light-front momentum.  This measures the difference between the light-front number-density of quarks with helicity parallel to that of the proton and the density of quarks with helicity antiparallel.  The connection between Eq.\,\eqref{isosymmetry} and helicity will not be surprising, given the relationship that may be drawn between the matrix structure $\gamma_5 \gamma_\mu$ and the Pauli-Lubanski four-vector; and it follows that
\begin{equation}
\rule{-1ex}{0ex} g_A = Z_A\! \int_0^1 \!dx\, \left[ \Delta u(x) + \Delta \bar u(x)- \Delta d(x) - \Delta \bar d(x)\right] ;  \label{gAhelicity}
\end{equation}
namely, the nonsinglet axial charge measures the difference in the light-front frame between the $u$- and $d$-quark contributions to the proton's helicity \cite{Jaffe:2000kr,Renner:2006sh}.  (Owing to the axial-vector Ward-Takahashi identity, the renormalisation constant for the axial-vector vertex satisfies $Z_A = Z_2$, with $Z_2$ discussed below.)

The induced pseudoscalar form factor, $G_P(q^2)$, holds its own fascinations, owing particularly to its connection with pion-nucleon interactions.  Fundamental to the character and strength of such interactions is dynamical chiral symmetry breaking (DCSB), the phenomenon responsible for both 98\% of the visible mass in the universe and masslessness of the chiral-limit pion \cite{Chang:2011vu}.  The existence of such a pion entails
\begin{equation}
\frac{q^2}{2 M_N}\, G_P(q^2) \stackrel{q^2 \sim 0}{=} 2 f^0_\pi g^0_{\pi N N}\,,
\end{equation}
where $f^0_\pi$ is the pion's leptonic decay constant and $g^0_{\pi N N}$ is the pion-nucleon coupling constant, where the superscript ``0'' indicates a quantity evaluated in the chiral limit.

Using a Gordon identity, chiral-limit axial-vector current conservation at the nucleon-level ($q_\mu\Lambda_{5\mu}^{pn}(p_f,p_i) = 0$) delivers the Goldberger-Treiman relation:
\begin{equation}
\label{GTrelation}
M_N^0\, g^0_A = f^0_\pi \, g^0_{\pi N N}\,.
\end{equation}
This identity has some curious implications.  For example, in the absence of DCSB, $f_\pi^0 = 0$ and hence no pseudoscalar meson couples to the weak interaction \cite{Holl:2004fr}.  It then follows from Eq.\,\eqref{GTrelation} that if a nucleon exists with a finite, nonzero mass in a universe without DCSB, $g_A^0=0$ for that nucleon; i.e., such nucleons, too, decouple from the weak interaction.  (We will subsequently return to this.)  In these circumstances then 
$g_A^0$ appears to serve as an order parameter for DCSB and a nonzero value of $g_A$ signals the presence of in-hadron quark condensates \cite{Brodsky:2008be,Brodsky:2009zd,Brodsky:2010xf,Chang:2011mu,Roberts:2011ea,Brodsky:2012ku}.

\section{Dressed-quarks}
Our goal is to elucidate a connection between $g_A$ and the strong physics of dressed-quarks, which are described in QCD by the gap equation:
\begin{eqnarray}
\lefteqn{
\nonumber S_f^{-1}(p) = Z_2 \,(i\gamma\cdot p + m_f^{\rm bm})}\\
&& + Z_1 \int^\Lambda_{dq}\!\! g^2 D_{\mu\nu}(p-q)\frac{\lambda^a}{2}\gamma_\mu S_f(q) \frac{\lambda^a}{2}\Gamma^f_\nu(q,p) ,
\label{gendseN}
\end{eqnarray}
where: $f$ denotes the quark's flavour; $D_{\mu\nu}$ is the gluon propagator; $\Gamma^f_\nu$, the quark-gluon vertex; $\int^\Lambda_{dq}$, a symbol representing a Poincar\'e invariant regularization of the four-dimensional integral, with $\Lambda$ the regularization mass-scale; $m_f^{\rm bm}(\Lambda)$, the current-quark bare mass; and $Z_{1,2}(\zeta^2,\Lambda^2)$, respectively, the vertex and quark wave-function renormalisation constants, with $\zeta$ the renormalisation point. 

The gap equation's solution is the dressed-quark propagator,
\begin{eqnarray}
S_f(p)
&=& 1/[i \gamma\cdot p \, A_f(p^2,\zeta^2) + B_f(p^2,\zeta^2)]\,,
\label{SgeneralNAB}\\
&=&Z_f(p^2,\zeta^2)/[i\gamma\cdot p + M_f(p^2)]\,.
\label{SgeneralN}
\end{eqnarray}
The mass function, $M_f(p^2)$, is independent of the renormalisation point; and the renormalised current-quark mass,
\begin{equation}
\label{mzeta}
m_f^\zeta = Z_m(\zeta,\Lambda) \, m_f^{\rm bm}(\Lambda) = Z_4^{-1} Z_2\, m_f^{\rm bm},
\end{equation}
wherein $Z_4$ is the renormalisation constant associated with the Lagrangian's mass-term.  The renormalisation-group invariant current-quark mass may be inferred via
\begin{equation}
\hat m_f = \lim_{p^2\to\infty} \left[\frac{1}{2}\ln \frac{p^2}{\Lambda^2_{\rm QCD}}\right]^{\gamma_m} M_f(p^2)\,,
\label{RGIcqmass}
\end{equation}
where $\gamma_m = 12/(33-2 N_f)$.  The chiral limit is
\begin{equation}
\hat m_f = 0\,.
\end{equation}

Chiral symmetry and its breaking pattern in QCD are expressed in the following axial-vector Ward-Takahashi identity:
\begin{eqnarray}
\nonumber
&& P_\mu \Gamma_{5\mu}^{fg}(k;P) + \, i\,[m_f(\zeta)+m_g(\zeta)] \,\Gamma_5^{fg}(k;P)\\
&=& S_f^{-1}(k_+) i \gamma_5 +  i \gamma_5 S_g^{-1}(k_-) \,,
\label{avwtimN}
\end{eqnarray}
where $\Gamma_{5\mu}^{fg}$ and $\Gamma_5^{fg}$ are, respectively, amputated axial-vector and pseudoscalar vertices.  They connect an outgoing quark of flavour $f$ and an incoming quark of flavour $g$, with total momentum $P=p_i+p_f$ and relative momentum $k=(1-\eta) p_i + \eta p_f$, where $\eta \in [0,1]$, and hence $k_+ = p_f = k + \eta P$, $k_- = p_i = k - (1-\eta) P$.  Owing to Poincar\'e covariance, no observable can legitimately depend on $\eta$; i.e., the definition of the relative momentum.
N.B.\ Equation~\eqref{avwtimN} is modified for flavourless pseudoscalar mesons and this leads to important differences in their behaviour, which are discussed in Ref.\,\cite{Bhagwat:2007ha}.

The vertices relevant to $\beta^-$-decay are $\Gamma_{5\mu}^{ud}$, $\Gamma_{5}^{ud}$ but with our assumption of isospin symmetry we can ignore the flavour labels and consider the diagonal elements $\Gamma_{5\mu}^{u=d}=\Gamma_{5\mu}$, $\Gamma_{5}^{u=d}=\Gamma_{5}$.  The axial-vector vertex then has the general form \cite{Maris:1997hd}
\begin{eqnarray}
\nonumber
\Gamma_{5\mu}(k;P) & = & \gamma_5 \left.[
\gamma_\mu F_R(k;P) 
+ k_\mu \gamma\cdot k G_R(k;P) \right. \\ 
&& \left. - \, \sigma_{\mu\nu} k_\nu H_R(k;P)  
\right]
+ \tilde \Gamma_{5\mu}(k;P) \nonumber \\
&& + \frac{P_\mu}{P^2+m_\pi^2}\,2\,f_\pi\, \Gamma_\pi(k;P)\,,
\label{genAVvtx}
\end{eqnarray}
where: $F_R$, $G_R$, $H_R$ and $\tilde \Gamma_{5\mu}(k;P)$ are regular in the neighbourhood of $[P^2+m_\pi^2]= 0$; $P_\mu \tilde \Gamma_{5\mu}(k;P) \sim P_\alpha P_\beta M_{\alpha\beta}(k;P)$, with $M_{\alpha\beta}(k;P)$ a matrix-valued function; and the pion's Bethe-Salpeter amplitude is
\begin{eqnarray}
\nonumber
\lefteqn{\Gamma_{\pi}(k;P) = \gamma_5
\left[ i E_{\pi}(k;P) + \gamma\cdot P F_{\pi}(k;P)  \right. }\\
&&  \left. + \, k\cdot P \gamma\cdot k \, G_{\pi}(k;P) + \sigma_{\mu\nu} k_\mu P_\nu H_{\pi}(k;P) \right]. 
\label{genGpi}
\end{eqnarray}
$\!$Combining now Eqs.\,\eqref{gendseN}, \eqref{SgeneralNAB}, \eqref{avwtimN}--\eqref{genGpi} and working in the chiral limit, one may derive \cite{Maris:1997hd} the following quark-level Goldberger-Treiman relations:
\begin{eqnarray}
\label{gtlrelE}
f_{\pi}^0 E_{\pi}(k;0) &=& B^0(k^2)\,,\\
\label{gtlrelF}
F_R(k;0) + 2 f_{\pi}^0 F_{\pi}(k;0) &=& A^0(k^2)\,,\\
\label{gtlrelG}
G_R(k;0) + 2 f_{\pi}^0 G_{\pi}(k;0) &=& \frac{d}{dk^2}A^0(k^2)\,,\\
\label{gtlrelH}
H_R(k;0) + 2 f_{\pi}^0 H_{\pi}(k;0) &=& 0\,.
\end{eqnarray}

These identities are of critical importance in QCD.
The first exposes the fascinating consequence that the solution of the two-body pseudoscalar bound-state problem is almost completely known once the one-body problem is solved for the dressed-quark propagator: the relative momentum within the bound-state is identified unambiguously with the momentum of the dressed-quark.  This last fact emphasises that Goldstone's theorem has a pointwise expression in QCD.  It is difficult to overestimate its importance for Standard Model physics.

The remaining three identities show that a pseudoscalar meson Goldstone boson \emph{must} contain components of pseudovector origin.  Some of the important corollaries of this result are exposed in Refs.\,\cite{Nguyen:2011jy,Roberts:2010rn,GutierrezGuerrero:2010md,Maris:1998hc,Roberts:2011wy}.
Herein, however, we reveal additional novel consequences of Eqs.\,\eqref{gtlrelF}--\eqref{gtlrelH}.

\section{Axial charge of a dressed-quark}
Consider the dressed-quark--axial-vector vertex, Eq.\,\eqref{genAVvtx}.  Only $F_R(k;P)$, the function associated with the Dirac structure $\gamma_5\gamma_\mu$, possesses an ultraviolet divergence in QCD perturbation theory.  Notably, in Landau gauge the renormalised amplitude
$F_R = 1$, up to next-to-leading-order perturbative corrections: one-loop corrections vanish.  (This may be derived following Ref.\,\cite{tarrach}.)
All other functions in the axial-vector vertex are power-law suppressed in the ultraviolet.  In perturbation theory, therefore, the quantity
\begin{equation}
g_A^q(k^2):=F_R(k;P=0)
\end{equation}
expresses the distribution of a current-quark's axial-charge.  It remains perturbatively close to unity.  (The impact of other components in Eq.\,\eqref{genAVvtx} is canvassed in Sec.\,\ref{sec:five}.  They do not materially affect our discussion.)

Nonperturbatively, however, the situation is very different, as may readily be illustrated. To this end, consider the symmetry-preserving regularisation of a vector$\,\times\,$vector contact-interaction detailed and employed in Refs.\,\cite{GutierrezGuerrero:2010md,Roberts:2010rn,Roberts:2011wy,Wilson:2011aa,Chen:2012qr}. As elucidated therein, in rainbow-ladder truncation\footnote{Rainbow-ladder is the leading-order in a systematic and symmetry-preserving truncation scheme for QCD's Dyson-Schwinger equations \cite{Munczek:1994zz,Bender:1996bb}.}
such an interaction produces results for low-momentum-transfer observables that are practically indistinguishable from those generated by more sophisticated interactions, such as that explained in Refs.\,\cite{Qin:2011dd,Qin:2011xq}.  The consequences of Eq.\,\eqref{gtlrelF} are dramatic in this context.
With the single parameter determining the interaction strength chosen small, $\alpha_{\rm IR}/\pi < 0.4$, then DCSB is absent and
\begin{equation}
g_{A_{\rm CN}}^q \stackrel{\mbox{\st{\scriptsize DCSB}}}{=}1\,,
\end{equation}
where ``\mbox{\footnotesize CN}'' denotes contact interaction.
On the other hand, with $\alpha_{\rm IR}/\pi \simeq 1$; namely, chosen commensurate with contemporary estimates of the zero-momentum value of a running-coupling in QCD \cite{Aguilar:2010gm,Boucaud:2010gr,Qin:2011dd,Boucaud:2011ug}, one obtains $A^0(k^2)=1$, $M^0(k^2) = M^0 =0.358\,$GeV, $M^0 F^0_\pi(k;0) = 0.46$, all $k$-independent with a contact interaction, and $f_\pi^0 = 0.1\,$GeV, so that
\begin{equation}
g_{A_{\rm CN}}^q = F_R^0(k;0) = 1 - 2 f_\pi^0\,F^0_\pi(k;0) = 0.74\,.
\label{gAcontact}
\end{equation}
Thus the quantity associated with the current-quark's axial-charge is markedly suppressed in the infrared owing to the nonperturbative phenomenon of DCSB.

To allay any concern that this outcome might be model specific, we compared Eq.\,\eqref{gAcontact} with the value produced by the most sophisticated rainbow-ladder interaction available \cite{Qin:2011dd}, which is detailed in App.\,\ref{subsec:RL}.  In this case one naturally finds a $k^2$-dependent form for $g_A^q$ and obtains $g_{A_{RL}}^q(k^2=0)= 0.81$ at a realistic value for the light-quark current-mass.

We are also able to compare these results with that produced by the most complete kernels available for the gap- and Bethe-Salpeter equations \cite{Chang:2011ei}.  These kernels, described briefly in App.\,\ref{subsec:DB} and denoted subsequently by ``\mbox{\footnotesize DB},'' incorporate essentially nonperturbative effects associated with DCSB, such as a dressed-quark anomalous magnetic moment \cite{Kochelev:1996pv,Diakonov:2002fq,Chang:2010hb}, and yield
\begin{equation}
g_{A_{DB}}^q(0)= 0.87 = 1.06\,g_{A_{RL}}^q(0)\,.
\label{AMMonRL}
\end{equation}
The infrared suppression is thus seen to be a generic feature of the axial-vector vertex.  Its impact is far-reaching since it will influence, e.g.: the leptonic radiative decays of charged light pseudoscalar mesons; and the nucleon's axial charge, as we shall subsequently see.

We stress that the infrared suppression of the axial-vector vertex contrasts markedly with the effect of dressing on the leading covariant, $\gamma_\mu$, in the vector vertex.  In this case the associated scalar function is bounded below by unity at $(k=0;P=0)$ owing to the vector Ward-Takahashi identity.  Indeed, with a momentum-dependent interaction this scalar function is always enhanced, as illustrated in Fig.\,2 of Ref.\,\cite{Chang:2011ei}.

At this point it is worth emphasising that Poincar\'e covariance demands that the general form for a pseudoscalar meson Bethe-Salpeter amplitude possess four components; namely, those appearing in Eq.\,\eqref{genGpi}.
Inspection of the Bethe-Salpeter equation for pseudoscalar mesons shows that a nonzero value for $E_\pi$ is the force behind $F_\pi \neq 0$, with the coupling fixed by the DCSB mass-scale, which is provided by the dressed-quark mass-function, $M$.  To be clear, $M\neq 0$ in the chiral limit entails $E_\pi\neq 0$, and together  these results require $F_\pi \neq 0$.  Readily apparent in the rainbow-ladder truncation, this is a general result, independent of the function chosen to represent the dressed-gluon and the \emph{Ansatz} for the dressed-quark-gluon vertex in the gap equation, Eq.\,\eqref{gendseN}.  For further confirmation, compare the results reported above with those in, e.g., Refs.\,\cite{Bender:1996bb,Maris:1997tm,Bender:2002as,Bhagwat:2004hn,GutierrezGuerrero:2010md}.

This explains why the appearance of $F_\pi\neq 0$ is a necessary consequence of DCSB.  Given an interaction with nontrivial momentum dependence, then $G_\pi$ and $H_\pi$ are also necessarily nonzero for the same reason.
Plainly, a complete expression of DCSB is not achieved merely by producing nonzero values for the in-pion condensate and pion leptonic decay constant.  The full structure of the Goldstone mode must also be described.
Finally, positivity of $f_\pi^0$ guarantees that of $F_\pi(k;0)$, and hence the second term on the left-hand-side of Eq.\,\eqref{gtlrelF} is positive.  This means that $F_R(k^2;P=0)$ is bounded above by $A^0(k^2)$ and approaches this function from below as $k^2\to \infty$.  (This is illustrated in Fig.\,8 of Ref.\,\cite{Maris:1997tm}.)  It does not, however guarantee $F_R(0;0)<1$.  That is a consequence of the dynamics which produces the Goldstone pion and sets the mass-scale for DCSB.
%



It is now \emph{a propos} to reconsider the role of $g_A$ in connection with DCSB.  In chiral-limit Dyson-Schwinger equation (DSE) studies, chiral symmetry restoration and deconfinement are coincident no matter which control parameter is varied.\footnote{See, e.g., the discussions in Refs.\,\cite{Roberts:2000aa,Bashir:2012fs}, for which it is important to note that light-quark confinement is not connected in any known manner with the static-quark potential.  It can instead be related to marked differences between the analytic properties of coloured and colour-singlet Schwinger functions \protect\cite{Krein:1990sf}.}
This supports a view that DCSB and confinement are intimately related; and we expect that in the presence of some agent which undermines the interaction strength required for DCSB, confinement is also lost.
Under these conditions $f_\pi^0=0$ and consequently $F_R(k^2;0) = A^0(k^2)$, following from Eq\,\eqref{gtlrelF}.
Should such circumstances correspond to a domain whereupon none of the interactions in the Standard Model is strong, then both functions will be unity up to perturbative corrections.
On the other hand, suppose that strong correlations remain after chiral symmetry restoration, such as may be in a putative strongly-coupled quark-gluon plasma, then $F_R(k^2;0) = A^0(k^2)>1$; i.e., both functions are actually enhanced above unity \cite{Roberts:2000aa,Bashir:2012fs}.
Evidently then a connection between the restoration of chiral symmetry and $g_A^0$ vanishing is not driven by changes at the level of the axial-vector dressed-quark vertex.

Consider now that the identity in Eq.\,\eqref{GTrelation} holds so long as chiral symmetry is dynamically broken and a nucleon exists with nonzero and finite mass, even under conditions that place the theory in the neighbourhood of $f_\pi^0=0^+$.  Given that $g_A^q \gtrsim 1$ in these circumstances, a vanishing of $g_A^0$ would require extraordinary and precise cancellations amongst the terms that constitute the nucleon's axial-charge matrix element; i.e., between the various contributions arising from the angular momentum correlations within the nucleon's Faddeev amplitude.  Owing to the power of symmetries in quantum field theory, this is conceivable but nevertheless improbable.
Given the preceding discussion it is more likely that the chiral-limit relationship $g_A^0\to 0$ is connected with dissolution of the nucleon bound-state at a point of coincident chiral symmetry restoration and deconfinement.  A realisation of this phenomenon is illustrated for the scalar and pseudoscalar meson sector in Sec.\,IV of Ref.\,\cite{Chang:2008ec}.  The conjecture may be tested using modern Faddeev equation treatments of the nucleon.

A vanishing of $g_A$ entails that the right-hand-side of Eq.\,\eqref{gAhelicity} is zero.  This expression is normally described as expressing the difference in the light-front frame between the $u$- and $d$-quark contributions to the proton's helicity.  How can that vanish? One is here considering the chiral limit.  Absent a DCSB mechanism, a chiral limit theory with massless quarks separates into two distinct, non-communicating theories: one for positive helicity states and another for negative helicity.  Each sub-theory has identical interactions and hence each will produce the same quark number distributions, labelled,  however, by opposite helicities.  Since there is no mechanism in the total theory that can flip helicity, the number of positive helicity states will always match the number with negative helicity.  Hence the result $g_A=0$ is achieved because each of the four terms in Eq.\,\eqref{gAhelicity} vanishes individually, irrespective of whether or not they are associated with a bound-state.

\section{Quark models and $\mathbf g_A$}
Related to constituent-quark model phenomenology, Eqs.\,\eqref{gAcontact}, \eqref{AMMonRL} are curious.  It is textbook knowledge (see, e.g., Ref.\,\cite{donoghue}) that constituent-quark models with spin-flavour wave-functions based on $SU(6)$ symmetry produce the following axial-charge of the nucleon:
\begin{eqnarray}
g_A &=& \frac{5}{3}\, g_A^Q \int d^3x\, \left[ u^2(x) - \frac{1}{3} v^2(x)\right]\\
    &=& \frac{5}{3}\, g_A^Q \left[ 1 - \frac{4}{3}\int d^3x\,v^2(x)\right]\,, \label{gAcorrelations}
\end{eqnarray}
where $g_A^Q$ is the axial-charge of a constituent-quark, and $u(x)$, $v(x)$ are, respectively, the upper and lower components of the nucleon's constituent-quark wave-function.
Plainly, in a nonrelativistic model, $v(x)\equiv 0$ and $g_A = (5/3) g_A^Q$, so that reproducing the empirical value of $g_A=1.27$ requires $g_{A_{\rm NR}}^Q=0.76$.  This value compares well with those in Eqs.\,\eqref{gAcontact}, \eqref{AMMonRL}.  Of course, the origin of the empirical value of $g_A$ is more complicated but nonperturbative dressing of $g_A^q$ plays a part.

A full explanation is suggested by Eq.\,\eqref{gAcorrelations}, which has two key features.  As we have described above, the first is dressing of the axial-vector vertex, an effect that  modifies the strength with which a dressed-quark couples to the $W$-boson.

The other is indicated by the second term within the parentheses in Eq.\,\eqref{gAcorrelations}:
\begin{equation}
{\mathpzc c}_v = \frac{4}{3}\int d^3x\,v^2(x)\,.
\end{equation}
This represents the appearance of $P$-wave quark orbital angular momentum in a relativistic constituent-quark model.

In a quantum field theory such as QCD, the nucleon is properly described by a Poincar\'e covariant Faddeev equation \cite{Cahill:1988dx}.  In this context, ${\mathpzc c}_v$ may be reinterpreted as signifying the impact of correlations within the nucleon's Faddeev wave-function, which possesses $S$-, $P$- and $D$-wave dressed-quark orbital angular momentum components in the nucleon's rest frame.  In the presence of DCSB, such correlations are strong.  For example, the $S$-wave-only contribution to the nucleon's normalisation is just 60\% \cite{Cloet:2007pi,Wilson:2011aa};\footnote{The canonical normalisation constant for the nucleon's Faddeev amplitude is equivalent to requiring that the nucleon's Dirac form factor is unity at zero momentum transfer.}
and it is known that altering the strength of quark orbital angular momentum correlations within the nucleon can materially affect $g_A$ \cite{Hecht:2001ry}.  Within this framework, therefore, the empirical value of $g_A$ embodies the outcome of interference between dressing the quark--$W$-boson vertex and angular momentum correlations within the nucleon's Faddeev amplitude.  This is not too surprising given the connection between $g_A$ and the $u$- and $d$-quark helicity distributions, expressed in Eq.\,\eqref{gAhelicity}.  Notably, however, the magnitude of the suppression of $g_A^q$ and the strength of orbital angular momentum correlations in bound-state wave functions are both driven by DCSB.

The latter is readily seen from Sec.\,III of Ref.\,\cite{Maris:2000ig}.  In the absence of DCSB, the amplitudes $F_\pi$, $G_\pi$, $H_\pi$ in Eq.\,\eqref{genGpi} vanish identically in the chiral limit, as does $M(p^2)$ in Eq.\,\eqref{SgeneralN}.  It follows that any correlation that survives is described by a Bethe-Salpeter wave-function: $\chi_\pi(k;P) = S(p) \Gamma_\pi(k;P) S(p) \propto \gamma_5$, a purely $S$-wave structure in rest-frame kinematics.

These causal relationships emphasise again, following identical conclusions drawn from computations of the pion and nucleon valence-quark distributions \cite{Hecht:2000xa,Roberts:2001tg,Holt:2010vj,Nguyen:2011jy,Wilson:2011aa,Chang:2012rk}, that understanding parton distribution functions (PDFs), as opposed to merely parametrising them, rests upon grasping the nature of DCSB in QCD.  It exposes the potential gains to be made in hadron physics by shifting theoretical focus from modelling PDFs to their well-constrained computation.  In this connection it should be borne in mind that the first three non-trivial PDF moments, which is the maximum that can be obtained from numerical simulations of lattice QCD \cite{Holt:2010vj}, are insufficient for PDF reconstruction: more than ten moments are required in order to constrain the large-$x$ exponent to better than 10\% \cite{RobertsPrivate}.

\section{Faddeev equation and $\mathbf g_A$}
\label{sec:five}
It is known that Faddeev equation models can be constructed to reproduce the empirical value of $g_A$ \cite{Oettel:2000jj,Roberts:2007jh}, unifying it in the process with other nucleon observables.  However, such studies employed axial-vector vertices that do not respect Eqs.\,\eqref{gtlrelF}--\eqref{gtlrelH}.  This is mended in Ref.\,\cite{Eichmann:2011pv}, which solves all elements of the problem -- the gap, Bethe-Salpeter and Faddeev equations -- in rainbow-ladder truncation.  That study, however, produces $g_{A_{RL}} = 0.99(2)$, underestimating the empirical value by 22\%.  Naturally, with $M_N^0$ and $f_\pi^0$ near to their experimental values, Eq.\,\eqref{GTrelation} entails that Ref.\,\cite{Eichmann:2011pv} underestimates $g_{\pi NN}$ by a similar amount.

The magnitude of the error is typical of rainbow-ladder truncation in those channels for which it is known and understood \emph{a priori} to be adequate.  In the sector of light-quark vector and flavour nonsinglet pseudoscalar mesons, over an illustrative basket of thirty-one calculated quantities tabulated in Ref.\,\cite{Maris:2003vk}, the truncation delivers a standard-deviation of 15\% in the relative error between experiment and theory \cite{Roberts:2007jh}.

Part of the remedy to this quantitative error lies in going beyond the leading-order truncation when solving the gap and Bethe-Salpeter equations.  This is now possible in a symmetry-preserving manner \cite{Chang:2009zb}, as indicated by Eq.\,\eqref{AMMonRL}.  Indeed, we have solved for the dressed-quark--axial-vector vertex using the kernels described in Ref.\,\cite{Chang:2011ei} and recapitulated in App.\,\ref{subsec:DB}, which are essentially nonperturbative, incorporating effects of DCSB that were not previously possible to express.  These kernels clarify a causal connection between DCSB and the splitting between vector and axial-vector mesons, and expose a key role played by the anomalous chromomagnetic moment of dressed-quarks \cite{Kochelev:1996pv,Bicudo:1998qb,Diakonov:2002fq,Ebert:2005es,Chang:2010hb} in determining the values of  observable quantities.

The general form of the transverse part of the axial-vector vertex, all that contributes directly to $g_A^q$, is
\begin{eqnarray}
%
%
\Gamma_{5\mu}^\perp(k;P) & = &
\gamma_{5}
\big[ \gamma_{\mu}^{\perp} F_{1}
-i\gamma_{\mu}^{\perp}\gamma\cdot \hat P k\cdot \hat P F_{2}
+ T_{\mu\nu} \sigma_{\nu\rho} k_\rho F_{3} \nonumber \\
&&
+[k_{\mu}^{\perp}\gamma\cdot \hat P + i \gamma_\mu^\perp \sigma_{\nu\rho} k_\nu \hat P_\rho] F_{4} -ik_{\mu}^{\perp} k\cdot \hat P F_{5} \nonumber\\
&& + k_{\mu}^{\perp}\gamma\cdot \hat P k\cdot \hat P F_{6}
+k_{\mu}^{\perp}\gamma\cdot k F_{7}
\nonumber\\
%
&& + k_{\mu}^{\perp}\sigma_{\nu\rho}k_\nu \hat P_\rho k\cdot \hat P F_{8}
\big], \label{AVtransverse}
\end{eqnarray}
where: $\{F_i | i=1,\ldots,8\}$ are scalar functions of $(k^2,\mbox{$k\cdot P$},P^2)$ that are even under $k\cdot P \to (- k\cdot P)$; $\hat P^2=1$; $T_{\mu\nu} = [\delta_{\mu\nu}-P_\mu P_\nu/P^2]$, $T_{\mu\nu}+L_{\mu\nu} = \delta_{\mu\nu}$; and $a_\mu^\perp = T_{\mu\nu} a_\nu$.  Since Eq.\,\eqref{AVtransverse} may simply be obtained from Eq.\,\eqref{genAVvtx} through contraction with $T_{\mu\nu}$, we have the following correspondences:
$F_1 \leftrightarrow F_R$, $F_7 \leftrightarrow G_R$, $F_3 \leftrightarrow H_R$.

\begin{figure}[t]
\centerline{%
\includegraphics[clip,width=0.45\textwidth]{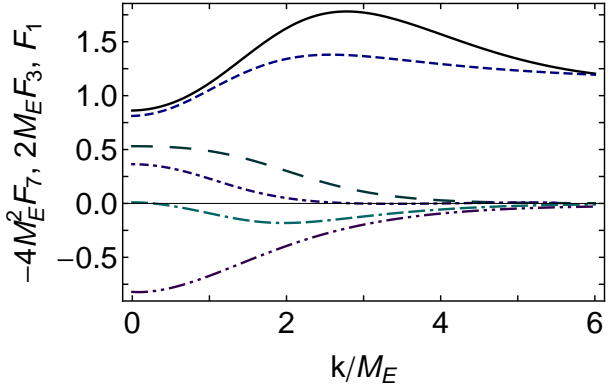}}
\caption{\label{FigF137} Selected functions in the axial-vector vertex, Eq.\,\protect\eqref{AVtransverse}: rainbow-ladder (RL) result cf.\ that obtained with DCSB-improved kernels for gap and Bethe-Salpeter equations (DB).
Curves, all dimensionless: \emph{solid} $F_1^{DB}$ and \emph{dashed} $F_1^{RL}$;
\emph{very-long-dashed} $2M_E^{DB} F_3^{DB}$ and \emph{dot-dashed} $2M_E^{RL} F_3^{RL}$;
and \emph{dot-dash-dash} $(-4M_E^{DB\,^2} F_7^{DB})$ and \emph{dot-dot-dash} $(-4M_E^{RL\,^2} F_7^{RL})$.
($M_E^{DB}=0.36\,$GeV and $M_E^{RL}=0.41$.)
}
\end{figure}

Given that a dressed-quark anomalous chromomagnetic moment produces a large dressed-quark anomalous magnetic moment \cite{Chang:2010hb}, one should at least expect that $F_3$, with its similar tensor structure, is significantly altered when proceeding beyond rainbow-ladder truncation.  In fact, all the scalar functions are materially modified on a domain $0 < |k|/M_E \lesssim 5$, where $M_E$ is the Euclidean constituent-quark mass, $\{M_E\}=\{\sqrt{s}\,|\,s=M^2(s),s>0\}$: $F_{1,2,3,5,6,8}$ magnitudes are enhanced, with $F_5$ also changing sign; and $F_{4,7}$ magnitudes are suppressed.  In Fig.\,\ref{FigF137} we illustrate the response of each of those functions appearing in Eqs.\,\eqref{gtlrelF}--\eqref{gtlrelH}.

With at least eight quantities reacting markedly to improvements in the DSE kernels, it is natural to seek a single measure that can illustrate the plausible consequences for $g_A$.  To this end we consider
\begin{equation}
\bar u(p_f) \Gamma_{5\mu}^\perp(k;P) u(p_i)\,, \label{uGu}
\end{equation}
where, at each value of $p^2>0$, the Euclidean spinors satisfy
$\gamma\cdot p \, u(p)= \varsigma_p \,u(p)$,
$\bar u(p) \gamma\cdot p = \bar u(p)\, \varsigma_p$, $\varsigma_p = M(p^2)$.  Focusing on the case $(k\cdot P=0$, $P^2=0)$ and using the appropriate Euclidean-Gordon identities, Eq.\,\eqref{uGu} yields an axial-charge distribution, which is complex for the on-Euclidean-mass-shell dressed-quarks:
\begin{equation}
g_A^{E_q}(k^2) = F_1(\varsigma_k^2;0) + i \varsigma_k \, F_3(\varsigma_k^2;0)\,.
\label{gAk2}
\end{equation}
With this kinematic arrangement, no other functions from Eq.\,\eqref{AVtransverse} contribute.  We reiterate and emphasise that Eq.\,\eqref{gAk2} is not an observable but rather an illustrative artifice: a simple quantitative measure of the impact of terms in Eq.\,\eqref{AVtransverse} on the infrared behaviour of the axial-charge of a dressed quark.

\begin{figure}[t]
\centerline{%
\includegraphics[clip,width=0.45\textwidth]{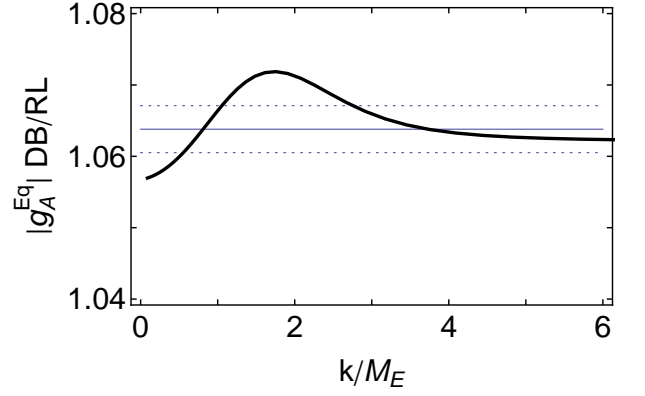}}
\caption{\label{ratioAbsgA} With the elements defined in association with Eq.\,\protect\eqref{gAk2}, the ratio $|g_A^{q DB}(k^2)|/|g_A^{q RL}(k^2)|$ (\emph{solid curve}).  The straight line with dotted outliers represents the band $1.064 \pm 0.003$.
}
\end{figure}

In Fig.\,\ref{ratioAbsgA} we plot the ratio $|g_A^{E_q DB}(k^2)|/|g_A^{E_q RL}(k^2)|$, which is the single measure we sought.  It assumes the value $1.064 \pm 0.003$, consistent with Eq.\,\eqref{AMMonRL}.  If one supposes that corrections to the Faddeev equation result for $g_A$, arising from bettering the rainbow-ladder computation of the dressed-quark axial-vector vertex, can simply be estimated by rescaling the axial-charge of a dressed-quark; viz., $g_A^{q RL} \to g_A^{q DB}=1.06 g_A^{q RL}$, then one infers a value of $g_A = 1.05(2)$ from Ref.\,\cite{Eichmann:2011pv}.  Consistent with studies of the nucleons' electromagnetic form factors \cite{Chang:2011tx}, this is an important but modest improvement.

The modest size of the improvement is good because the utility of rainbow-ladder truncation would have been much reduced if the magnification were too large.  Notwithstanding this, the constant rescaling probably underestimates the effect, given the structure apparent in Fig.\,\ref{FigF137}.  One would better reckon the correction by building an \emph{Ansatz} for $\Gamma_{5\mu}$, consistent with the algebraic constraints and numerical results we have elucidated, and employing that in a Faddeev equation computation of the nucleon's weak and strong form factors.  One should bear in mind, however, that correcting the gap and Bethe-Salpeter equation kernels is not the complete picture.  The Faddeev equation kernel and associated interaction current should also be modified.  These modifications, too, must affect $g_A$. 

In association with this we note that our dressed kernels do not contain pieces that might reasonably be described as corresponding to meson-cloud effects.  Considered analysis of such contributions is expected to further increase the value of $g_A$ by $\lesssim 10$\% \protect\cite{Thomas:1981vc,Schreiber:1988uw}.  They can be added to our kernels and interaction current without concern for overcounting and hence their effect may also be explored.
The role played by a meson-cloud in forming $g_A$ is, in fact, much discussed.  A contemporary effective field theory perspective may be traced from Ref.\,\protect\cite{Schindler:2006jq,Bernard:2007zu}; that within lattice-QCD from Refs.\,\protect\cite{Renner:2010ks,Wittig:2012ha,Hall:2012qn}; and that within models of nucleon structure from Refs.\,\protect\cite{Bass:2009ed,Bijker:2009up}.
%

Improvement of the gap and Bethe-Salpeter kernels should also affect the result for the nucleon's axial radius, $r_A$, quite simply because the rainbow-ladder truncation is unable to explain the location of the $a_1$-meson pole in the axial-vector vertex, whereas the DCSB-corrected kernels, App.\,\ref{subsec:DB}, resolve this longstanding problem \cite{Chang:2011ei}.
In underestimating the mass of the $a_1$ meson, the rainbow-ladder truncation overestimates the contribution to $r_A$ from the associated pole.  It is likely, therefore, to overestimate this radius or, equally, understate the mass-scale, $m_A$, that characterises evolution of the nucleon's axial form factor in the neighbourhood of $P^2=0$.

\begin{figure}[t]
\centerline{%
\includegraphics[clip,width=0.45\textwidth]{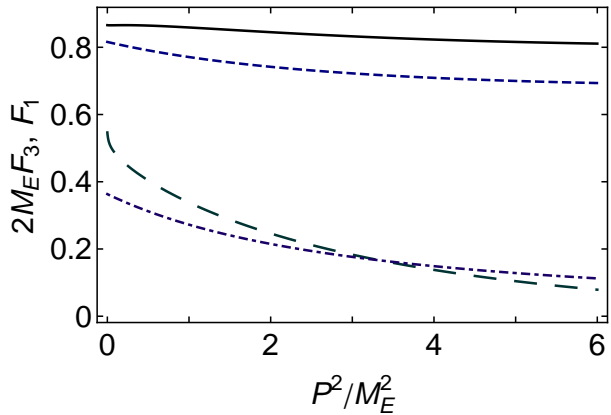}}
\caption{\label{F1F3Q2} $P^2$ dependence of the functions appearing in Eq.\,\protect\eqref{gAk2}.  Curves, all dimensionless: \emph{solid} $F_1^{DB}(k^2=0;P^2)$ and \emph{dashed} $F_1^{RL}(k^2=0;P^2)$; and \emph{long-dashed} $2 M_E^{DB} F_3^{DB}(k^2=0;P^2)$ and \emph{dot-dashed} $2 M_E^{RL} F_3^{RL}(k^2=0;P^2)$.
($M_E^{DB}=0.36\,$GeV and $M_E^{RL}=0.41$.)
}
\end{figure}

In Fig.\,\ref{F1F3Q2} we depict the $P^2$ dependence of the functions in Eq.\,\eqref{gAk2}.  Once again, the effects of improving the kernels are measurable.  An indication of its impact is the ratio of mass-scales that characterise monopole fits to the $F_1$ functions in the figure: $\sigma^{DB}/\sigma^{RL} = 1.65$.  This matches well with the expectation just described.  It is curious, however, because Ref.\,\cite{Eichmann:2011pv} reports $m_A^{RL}=1.28(6)$, which is already at the upper limit of values inferred from experiment (reviewed in Ref.\,\cite{Bernard:2001rs}): $m_A \in [1.0,1.3]$.  This corresponds to a value of $r_A$ at the lower limit of experiment.  It is relevant here to note that corrections to the Faddeev kernel and associated interaction current can plausibly magnify correlations within the nucleon and their impact on interactions, as the discussion of Fig.\,\ref{FigF137} shows they do for quark-antiquark systems.  Such effects would serve to increase $r_A$.





\section{Epilogue}
The nucleon's axial charge, $g_A$, expresses features that are both fundamental to the strong interaction and crucial to its connection with weak interaction physics.  It is thus important to understand how its strength originates within QCD and is thereby connected with dynamical chiral symmetry breaking (DCSB), the source of more than 98\% of visible mass in the Universe.

In pursuing this goal we demonstrated that DCSB suppresses the axial-charge distribution of a dressed-quark, $g_A^q$, at momenta $k \lesssim M_E$, where $M_E\sim 0.4\,$GeV is the mass-scale associated with DCSB.  Conversely, quark-level Goldberger-Treiman relations indicate that $g_A^q \simeq 1$ in the absence of DCSB.  This result, combined with the nucleon's Goldberger-Treiman relation, led us to a view that $g_A$ vanishes with the restoration of chiral symmetry because no nucleon bound-state survives the associated transition.

Consistent with inferences based on constituent-quark models, we found a suppression of $g_A^q$ to be part of an explanation for the value of the nucleon's axial charge.  Critical, too, however, is the presence of dressed-quark angular momentum correlations in the nucleon's rest-frame Faddeev wave-function, and hence in (almost) every frame as a result of Poincar\'e covariance.  (It would be an exceptional Poincar\'e transformation that transferred an observer to a frame in which every vestige of orbital angular momentum was eliminated.)

The Poincar\'e covariant Faddeev equation is a natural means by which to describe the structure of the nucleon bound-state.  As one of QCD's Dyson-Schwinger equations (DSEs), it is a critical element in a unified symmetry-preserving explanation of meson and baryon properties.  At leading-order in the most commonly used truncation (namely, rainbow-ladder approximation), this approach produces a value of $g_A$ that is $22$\% smaller than experiment.  We explained that this is a good result given the established level of accuracy that one may expect at leading-order

Complementing this, we argued that well-constrained improvements to the kernels of the gap and Bethe-Salpeter equations, which incorporate essentially nonperturbative corrections to the rainbow-ladder truncation, increase $g_A^q$ and are therefore likely to improve the DSE result for $g_A$.  Such corrections also affect the nucleon's axial radius.  In both cases we saw that agreement with experiment will require similar improvements to the Faddeev kernel and interaction current.

In closing it is worth reiterating that one should generally expect a $\sim 15$\% mismatch between experiment and results obtained in the internally consistent application of rainbow-ladder truncation in those channels for which the truncation is most reliable; namely, ground-state light-quark vector and flavour nonsinglet pseudoscalar mesons, and the nucleon and $\Delta$ ground-states.  Uniformly precise agreement would indicate serious deficiencies in the method -- a misuse of degrees-of-freedom, for example -- and diminish materially its capacity to provide insights into strong QCD.

The truncation's simplicity is a strength.  One can tolerate such modest disagreement with experiment when the result is ready computation of a diverse array of phenomena, their semiquantitative connection with fundamental elements in QCD, and enabling of a clear identification of familial relationships between them.

Notwithstanding these features, improvement is now possible and necessary.  Fuller incorporation of DCSB into bound-state kernels and interaction currents will enable the better informed feedback between experiment and theory that is necessary to understanding: confinement and DCSB in QCD; the nature of connections between them; and how they affect observables ranging from elastic and transition form factors to parton distribution functions.

\section*{Acknowledgments}
%
We are grateful for valuable input from A.~Bashir, I.\,C.~Clo\"et, B.~El-Bennich, P.\,C.~Tandy and D.\,J.~Wilson.
This work was supported by:
U.\,S.\ Department of Energy, Office of Nuclear Physics, contract no.~DE-AC02-06CH11357;
and Forschungszentrum J\"ulich GmbH.

\appendix
\section{Interaction kernels}
\label{sec:kernels}
\subsection{Rainbow-ladder}
\label{subsec:RL}
In a sophisticated rainbow-ladder DSE study the model input is expressed in a statement about the nature of the gap equation's kernel at infrared momenta, since the behaviour at momenta $k^2\gtrsim 2\,$GeV$^2$ is fixed by perturbation theory and the renormalisation group \cite{Jain:1993qh,Maris:1997tm}.  In Eq.\,\eqref{gendseN}, this amounts to writing ($k=p-q$)
\begin{eqnarray}
\nonumber
\lefteqn{
Z_1 g^2 D_{\mu\nu}(k) \Gamma_\nu(q,p) = k^2 {\cal G}(k^2)
D^{\rm free}_{\mu\nu}(k) \gamma_\nu}\\
&=&   \left[ k^2 {\cal G}_{\rm IR}(k^2) + 4\pi \tilde\alpha_{\rm pQCD}(k^2) \right]
D^{\rm free}_{\mu\nu}(k) \gamma_\nu ,
\label{rainbowdse}
\end{eqnarray}
wherein $D^{\rm free}_{\mu\nu}(k)$ is the Landau-gauge free-gauge-boson propagator;
$\tilde\alpha_{\rm pQCD}(k^2)$ is a bounded, monotonically-decreasing regular continuation of the perturbative-QCD running coupling to all values of spacelike-$k^2$; and ${\cal G}_{\rm IR}(k^2)$ is an \emph{Ansatz} for the interaction at infrared momenta: ${\cal G}_{\rm IR}(k^2)\ll \tilde\alpha_{\rm pQCD}(k^2)$ $\forall k^2\gtrsim 2\,$GeV$^2$.  The form of ${\cal G}_{\rm IR}(k^2)$ determines whether confinement and/or DCSB are realised in solutions of the gap equation.  

The interaction in Ref.\,\cite{Qin:2011dd} is
\begin{equation}
\label{CalGQC}
{\cal G}(s) = \frac{8 \pi^2}{\omega^4} D \, {\rm e}^{-s/\omega^2}
+ \frac{8 \pi^2 \gamma_m\, {\cal F}(s)}{\ln [ \tau + (1+s/\Lambda_{\rm QCD}^2)^2]} ,
\end{equation}
where: $\gamma_m = 12/(33-2 N_f)$, $N_f=4$, $\Lambda_{\rm QCD}=0.234\,$GeV; $\tau={\rm e}^2-1$; and ${\cal F}(s) = \{1 - \exp(-s/[4 m_t^2])\}/s$, $m_t=0.5\,$GeV.
With $D\omega = \,$constant, light-quark observables are independent of the value of $\omega \in [0.4,0.6]\,$GeV.  We used $D\omega = (0.8\,{\rm GeV})^3$ and $\omega =0.5\,$GeV.

In the rainbow-ladder truncation, in the isospin symmetric limit, the inhomogeneous axial-vector Bethe-Salpeter equation is
\begin{eqnarray}
\nonumber
\lefteqn{\Gamma_{5\mu}(k;P) = Z_2 \gamma_5\gamma_\mu }\\
&&
 - \frac{4}{3}\int_{dq}^\Lambda {\cal G}(k-q)\, D^{\rm free}_{\mu\nu}(k-q)\, \gamma_\alpha  \chi_{5\mu}(q;P) \gamma_\beta\,,
\end{eqnarray}
where $\chi_{5\mu} = S(q_+) \Gamma_{5\mu}^{fg}(q;P) S(q_-)$.

Regarding renormalisation of the gap and inhomogeneous Bethe-Salpeter equations, we follow precisely the procedures of Refs.\,\cite{Maris:1997tm,Maris:1999nt} and use the same renormalisation point; i.e., $\zeta=19\,$GeV.  A current-quark mass of $m^\zeta=3.4\,$MeV produces $m_\pi=0.136\,$GeV.

\subsection{DCSB-improved kernel}
\label{subsec:DB}
The DCSB-improved kernel is specified by a dressed-quark-gluon vertex and a Bethe-Salpeter kernel determined therefrom.

In the gap equation, Eq.\,\eqref{gendseN}, we use \cite{Chang:2011ei}
\begin{equation}
Z_1 g^2 D_{\rho \sigma}(t) \Gamma_\sigma(q,q+t)
= {\cal G}(t^2) \, D_{\rho\sigma}^{\rm free}(t) Z_2 \tilde\Gamma_\sigma(q,q+t)\,, \label{KernelAnsatz}
\end{equation}
with ${\cal G}$ from Eq.\,\eqref{CalGQC}, $\omega=0.5\,$GeV but $D\omega = (0.52\,{\rm GeV})^3$, a change required to ensure the dressed-kernels produce physical observables which match those obtained in rainbow-ladder truncation when that is reliable;
\begin{eqnarray}
\label{ourvtx}
\tilde\Gamma_\mu(p_1,p_2)  & = & \Gamma_\mu^{\rm BC}(p_1,p_2) +
\Gamma_\mu^{\rm acm}(p_1,p_2)\,;\\
\nonumber
i\Gamma_\mu^{\rm BC}(p_1,p_2)  & = &
i\Sigma_A(p_1^2,p_2^2)\,\gamma_\mu + 2 \ell_\mu \left[ i\gamma\cdot \ell \,\Delta_A(p_1^2,p_2^2)  \right. \\
&&  \left. + \Delta_B(p_1^2,p_2^2)\right] ,
\label{bcvtx}
\end{eqnarray}
where the first term was introduced in Ref.\,\cite{Ball:1980ay}, with $\Sigma_{\phi}(p_1^2,p_2^2) = [\phi(p_1^2)+\phi(p_2^2)]/2$, $\Delta_{\phi}(p_1^2,p_2^2) = [\phi(p_1^2)-\phi(p_2^2)]/[p_1^2-p_2^2]$, $2 \ell = p_1+p_2$; and the anomalous chromomagnetic moment piece is \cite{Chang:2010hb}
\begin{equation}
\Gamma_\mu^{\rm acm}(p_1,p_2) = \Gamma_\mu^{\rm acm_4}(p_1,p_2) + \Gamma_\mu^{\rm acm_5}(p_1,p_2)\,,
\end{equation}
with ($k=p_1-p_2$, $T_{\mu\nu} = \delta_{\mu\nu} - k_\mu k_\nu/k^2$, $a_\mu^{\rm T} := T_{\mu\nu}a_\nu$)
\begin{eqnarray}
\Gamma_\mu^{\rm acm_4} &=& [ \ell_\mu^{\rm T} \gamma\cdot  k + i \gamma_\mu^{\rm T} \sigma_{\nu\rho}\ell_\nu k_\rho] \tau_4(p_1,p_2)\,,\\
%
%
\Gamma_\mu^{\rm acm_5} & =& \sigma_{\mu\nu}k_\nu\tau_5(p_1,p_2)\,,\\
\tau_4 &=& \frac{2 \tau_5(p_1,p_2)}{\mathcal{M}(p_1^2,p_2^2)}\,,
\label{tau4}
\end{eqnarray}
$\tau_5 =  \eta\, \Delta_B(p_1^2,p_2^2)$, $\eta=0.65$ \cite{Chang:2011ei}; and ${\cal M}(x,y)=[x+M(x)^2+y+M(y)^2]/(2[M(x)+M(y)])$.
%

The inhomogeneous Bethe-Salpeter equation is
\begin{eqnarray}
\nonumber
\Gamma_{5\mu}(k;P) & = & Z_2 \gamma_5\gamma_\mu \\
&& \nonumber
- Z_2\!\int_{dq} {\cal G}(k-q) \, D_{\rho\sigma}^{\rm free}(k-q)\frac{\lambda^a}{2}\,\gamma_\alpha S(q_+) \\
\nonumber
&&  \times \Gamma_{5\mu}(q;P) S(q_-) \frac{\lambda^a}{2}\,\tilde\Gamma_\beta(q_-,k_-) \\
\nonumber
&& + Z_1 \! \int_{dq} g^2D_{\alpha\beta}(k-q)\, \frac{\lambda^a}{2}\,\gamma_\alpha S_f(q_+)\\
&& \times  \frac{\lambda^a}{2} \Lambda_{5\mu\beta}(k,q;P), \label{genbse}
\end{eqnarray}
where the four-point function $\Lambda_{5\mu\beta}$ is completely defined \cite{Chang:2009zb} via the quark self-energy and hence the quark-gluon vertex, $\Gamma_\mu$.  Crucially, $\Lambda_{5\mu\beta}$ satisfies a Ward-Takahashi identity \cite{Chang:2009zb}, whose solution provides a symmetry-preserving \emph{Ansatz} consistent with $\Gamma_\mu$.  We use
\begin{eqnarray}
\nonumber
2 \Lambda_{5\beta(\mu)} &= & [\tilde \Gamma_{\beta}(q_{+},k_{+})+\gamma_{5}\tilde \Gamma_{\beta}(q_{-},k_{-})
\gamma_{5}]\\
&& \times \frac{1}{S^{-1}(k_{+})+S^{-1}(-k_{-})}\Gamma_{5(\mu)}(k;P)\nonumber\\
\nonumber
&+&\Gamma_{5(\mu)}(q;P)\frac{1}{S^{-1}(-q_{+})+S^{-1}(q_{-})}\\
&& \times [\gamma_{5}\tilde\Gamma_{\beta}(q_{+},k_{+})\gamma_{5}
+\tilde\Gamma_{\beta}(q_{-},k_{-})].
\end{eqnarray}

Regarding renormalisation, here, too, we follow the procedures of Refs.\,\cite{Maris:1997tm,Maris:1999nt} and use the same renormalisation point; i.e., $\zeta=19\,$GeV.  A current-quark mass of $m^\zeta=3.7\,$MeV produces $m_\pi=0.138\,$GeV.

\bibliographystyle{../../../../../zProc/z10KITPC/h-physrev4}
\bibliography{../../../../../CollectedBiB}

\begin{thebibliography}{10}

\bibitem{Fermi:1934sk}
E.~Fermi,
\newblock Nuovo Cim. {\bf 11}, 1 (1934).

\bibitem{Fermi:1934hr}
E.~Fermi,
\newblock Z. Phys. {\bf 88}, 161 (1934).

\bibitem{Nakamura:2010zzi}
K.~Nakamura {\em et~al.},
\newblock J. Phys. {\bf G37}, 075021 (2010).

\bibitem{Edwards:2005ym}
R.~Edwards {\em et~al.},
\newblock Phys.Rev.Lett. {\bf 96}, 052001 (2006).

\bibitem{Renner:2006sh}
D.~B. Renner {\em et~al.},
\newblock J. Phys. Conf. Ser. {\bf 46}, 152 (2006).

\bibitem{Roberts:2007jh}
C.~D. Roberts, M.~S. Bhagwat, A.~H{\"o}ll and S.~V. Wright,
\newblock Eur. Phys. J. ST {\bf 140}, 53 (2007).

\bibitem{Yamazaki:2008py}
T.~Yamazaki {\em et~al.},
\newblock Phys. Rev. Lett. {\bf 100}, 171602 (2008).

\bibitem{Eichmann:2011pv}
G.~Eichmann and C.~Fischer,
\newblock Eur. Phys. J. {\bf A48}, 9 (2012).

\bibitem{Renner:2010ks}
D.~B. Renner,
\newblock PoS {\bf LAT2009}, 018 (2009).

\bibitem{Wittig:2012ha}
H.~Wittig,
\newblock PoS {\bf LATTICE2011}, 025 (2011).

\bibitem{Hall:2012qn}
N.~Hall, A.~Thomas, R.~Young and J.~Zanotti,
\newblock arXiv:1205.1608 [hep-lat], \emph{Volume Dependence of the Axial
  Charge of the Nucleon}.

\bibitem{Schindler:2006jq}
M.~Schindler and S.~Scherer,
\newblock Eur. Phys. J. {\bf A32}, 429 (2007).

\bibitem{Jaffe:2000kr}
R.~Jaffe,
\newblock Phil. Trans. Roy. Soc. Lond. {\bf A359}, 391 (2001).

\bibitem{Chang:2011vu}
L.~Chang, C.~D. Roberts and P.~C. Tandy,
\newblock Chin. J. Phys. {\bf 49}, 955 (2011).

\bibitem{Holl:2004fr}
A.~H{\"o}ll, A.~Krassnigg and C.~D. Roberts,
\newblock Phys. Rev. {\bf C70}, 042203 (2004).

\bibitem{Brodsky:2008be}
S.~J. Brodsky and R.~Shrock,
\newblock Phys. Lett. {\bf B666}, 95 (2008).

\bibitem{Brodsky:2009zd}
S.~J. Brodsky and R.~Shrock,
\newblock Proc. Nat. Acad. Sci. {\bf 108}, 45 (2011), [See also S.\,J.~Brodsky
  and R.~Shrock, arXiv:0803.2541, arXiv:0803.2554].

\bibitem{Brodsky:2010xf}
S.~J. Brodsky, C.~D. Roberts, R.~Shrock and P.~C. Tandy,
\newblock Phys. Rev. {\bf C82}, 022201(R) (2010).

\bibitem{Chang:2011mu}
L.~Chang, C.~D. Roberts and P.~C. Tandy,
\newblock Phys. Rev. {\bf C85}, 012201(R) (2012).

\bibitem{Roberts:2011ea}
C.~D. Roberts,
\newblock Few Body Syst. {\bf 52}, 345 (2012).

\bibitem{Brodsky:2012ku}
S.~J. Brodsky, C.~D. Roberts, R.~Shrock and P.~C. Tandy,
\newblock Phys. Rev. {\bf C{\,}85}, 065202 (2012).

\bibitem{Bhagwat:2007ha}
M.~S. Bhagwat, L.~Chang, Y.-X. Liu, C.~D. Roberts and P.~C. Tandy,
\newblock Phys. Rev. {\bf C76}, 045203 (2007).

\bibitem{Maris:1997hd}
P.~Maris, C.~D. Roberts and P.~C. Tandy,
\newblock Phys. Lett. {\bf B420}, 267 (1998).

\bibitem{Nguyen:2011jy}
T.~Nguyen, A.~Bashir, C.~D. Roberts and P.~C. Tandy,
\newblock Phys. Rev. {\bf C83}, 062201(R) (2011).

\bibitem{Roberts:2010rn}
H.~L.~L. Roberts, C.~D. Roberts, A.~Bashir, L.~X. Guti{\'e}rrez-Guerrero and
  P.~C. Tandy,
\newblock Phys. Rev. {\bf C82}, 065202 (2010).

\bibitem{GutierrezGuerrero:2010md}
L.~X. Guti{\'e}rrez-Guerrero, A.~Bashir, I.~C. Clo{\"e}t and C.~D. Roberts,
\newblock Phys. Rev. {\bf C81}, 065202 (2010).

\bibitem{Maris:1998hc}
P.~Maris and C.~D. Roberts,
\newblock Phys. Rev. {\bf C58}, 3659 (1998).

\bibitem{Roberts:2011wy}
H.~L.~L. Roberts, A.~Bashir, L.~X. Guti{\'e}rrez-Guerrero, C.~D. Roberts and
  D.~J. Wilson,
\newblock Phys. Rev. {\bf C83}, 065206 (2011).

\bibitem{tarrach}
P.~Pascual and R.~Tarrach,
\newblock {\em QCD: Renormalization for the Practitioner} (Springer-Verlag,
  Berlin, 1984),
\newblock {L}ecture Notes in Physics \textbf{194}.

\bibitem{Wilson:2011aa}
D.~J. Wilson, I.~C. Clo{\"e}t, L.~Chang and C.~D. Roberts,
\newblock Phys. Rev. {\bf C85}, 025205 (2012).

\bibitem{Chen:2012qr}
C.~Chen, L.~Chang, C.~D. Roberts, S.~Wan and D.~J. Wilson,
\newblock Few Body Syst. \emph{in press}  (2012), [arXiv:1204.2553 nucl-th].

\bibitem{Munczek:1994zz}
H.~J. Munczek,
\newblock Phys. Rev. {\bf D52}, 4736 (1995).

\bibitem{Bender:1996bb}
A.~Bender, C.~D. Roberts and L.~von Smekal,
\newblock Phys. Lett. {\bf B380}, 7 (1996).

\bibitem{Qin:2011dd}
S.-x. Qin, L.~Chang, Y.-x. Liu, C.~D. Roberts and D.~J. Wilson,
\newblock Phys. Rev. {\bf C84}, 042202(R) (2011).

\bibitem{Qin:2011xq}
S.-x. Qin, L.~Chang, Y.-x. Liu, C.~D. Roberts and D.~J. Wilson,
\newblock Phys. Rev. {\bf C85}, 035202 (2012).

\bibitem{Aguilar:2010gm}
A.~C. Aguilar, D.~Binosi and J.~Papavassiliou,
\newblock JHEP {\bf 07}, 002 (2010).

\bibitem{Boucaud:2010gr}
P.~Boucaud {\em et~al.},
\newblock Phys. Rev. {\bf D82}, 054007 (2010).

\bibitem{Boucaud:2011ug}
P.~Boucaud {\em et~al.},
\newblock Few Body Syst. \emph{in press}  (2012).

\bibitem{Chang:2011ei}
L.~Chang and C.~D. Roberts,
\newblock Phys. Rev. {\bf C85}, 052201(R) (2012).

\bibitem{Kochelev:1996pv}
N.~I. Kochelev,
\newblock Phys. Lett. {\bf B426}, 149 (1998).

\bibitem{Diakonov:2002fq}
D.~Diakonov,
\newblock Prog. Part. Nucl. Phys. {\bf 51}, 173 (2003).

\bibitem{Chang:2010hb}
L.~Chang, Y.-X. Liu and C.~D. Roberts,
\newblock Phys. Rev. Lett. {\bf 106}, 072001 (2011).

\bibitem{Maris:1997tm}
P.~Maris and C.~D. Roberts,
\newblock Phys. Rev. {\bf C56}, 3369 (1997).

\bibitem{Bender:2002as}
A.~Bender, W.~Detmold, C.~D. Roberts and A.~W. Thomas,
\newblock Phys. Rev. {\bf C65}, 065203 (2002).

\bibitem{Bhagwat:2004hn}
M.~S. Bhagwat, A.~H{\"o}ll, A.~Krassnigg, C.~D. Roberts and P.~C. Tandy,
\newblock Phys. Rev. {\bf C70}, 035205 (2004).

\bibitem{Roberts:2000aa}
C.~D. Roberts and S.~M. Schmidt,
\newblock Prog. Part. Nucl. Phys. {\bf 45}, S1 (2000).

\bibitem{Bashir:2012fs}
A.~Bashir {\em et~al.},
\newblock Commun. Theor. Phys. {\bf 58}, 79 (2012).

\bibitem{Krein:1990sf}
C.~D. Roberts, A.~G. Williams and G.~Krein,
\newblock Int. J. Mod. Phys. {\bf A7}, 5607 (1992).

\bibitem{Chang:2008ec}
L.~Chang {\em et~al.},
\newblock Phys. Rev. {\bf C79}, 035209 (2009).

\bibitem{donoghue}
J.~F. Donoghue, E.~Golowich and B.~R. Holstein,
\newblock {\em Dynamics of the Standard Model} (University Press, Cambridge,
  1994),
\newblock {C}ambridge Monographs on Particle Physics, Nuclear Physics and
  Cosmology \textbf{2}.

\bibitem{Cahill:1988dx}
R.~T. Cahill, C.~D. Roberts and J.~Praschifka,
\newblock Austral. J. Phys. {\bf 42}, 129 (1989).

\bibitem{Cloet:2007pi}
I.~C. Clo{\"e}t, A.~Krassnigg and C.~D. Roberts,
\newblock (arXiv:0710.5746 [nucl-th]),
\newblock In {\it Proceedings} {\it of} \emph{11th} \emph{International}
  \emph{Conference} \emph{on} {Meson-Nucleon Physics and} {\it the} {\it
  Structure} \emph{of} \emph{the} \emph{Nucleon} \emph{(MENU 2007)},
  J{\"u}lich, Germany, 10-14 Sep 2007, eds.\ H.~Machner and S.~Krewald, paper
  125.

\bibitem{Hecht:2001ry}
M.~B. Hecht, C.~D. Roberts and S.~M. Schmidt,
\newblock Phys. Rev. {\bf C64}, 025204 (2001).

\bibitem{Maris:2000ig}
P.~Maris, C.~D. Roberts, S.~M. Schmidt and P.~C. Tandy,
\newblock Phys. Rev. {\bf C63}, 025202 (2001).

\bibitem{Hecht:2000xa}
M.~B. Hecht, C.~D. Roberts and S.~M. Schmidt,
\newblock Phys. Rev. {\bf C63}, 025213 (2001).

\bibitem{Roberts:2001tg}
C.~D. Roberts,
\newblock Nucl. Phys. Proc. Suppl. {\bf 108}, 227 (2002).

\bibitem{Holt:2010vj}
R.~J. Holt and C.~D. Roberts,
\newblock Rev. Mod. Phys. {\bf 82}, 2991 (2010).

\bibitem{Chang:2012rk}
L.~Chang, C.~D. Roberts and D.~J. Wilson,
\newblock PoS {\bf QCD-TNT-II}, 039 (2012).

\bibitem{RobertsPrivate}
C.~D. Roberts,
\newblock Deconstructing \mbox{QCD}'s \mbox{Goldstone Modes},
\newblock
  \url{http://www.phy.anl.gov/ectdrell-yan/1stDAY/1205ECTCDRoberts.pptx},
\newblock Presentation at the ECT$^\ast$ Workshop on \emph{Drell-Yan Scattering
  and the Structure of Hadrons}, Trento, Italy, 21-25 May 2012.

\bibitem{Oettel:2000jj}
M.~Oettel, R.~Alkofer and L.~von Smekal,
\newblock Eur.Phys.J. {\bf A8}, 553 (2000).

\bibitem{Maris:2003vk}
P.~Maris and C.~D. Roberts,
\newblock Int. J. Mod. Phys. {\bf E12}, 297 (2003).

\bibitem{Chang:2009zb}
L.~Chang and C.~D. Roberts,
\newblock Phys. Rev. Lett. {\bf 103}, 081601 (2009).

\bibitem{Bicudo:1998qb}
P.~J.~A. Bicudo, J.~E. F.~T. Ribeiro and R.~Fernandes,
\newblock Phys. Rev. {\bf C59}, 1107 (1999).

\bibitem{Ebert:2005es}
D.~Ebert, R.~N. Faustov and V.~O. Galkin,
\newblock Eur. Phys. J. {\bf C47}, 745 (2006).

\bibitem{Chang:2011tx}
L.~Chang, I.~C. Clo{\"e}t, C.~D. Roberts and H.~L.~L. Roberts,
\newblock AIP Conf. Proc. {\bf 1354}, 110 (2011).

\bibitem{Thomas:1981vc}
A.~W. Thomas, S.~Theberge and G.~A. Miller,
\newblock Phys. Rev. {\bf D24}, 216 (1981).

\bibitem{Schreiber:1988uw}
A.~W. Schreiber and A.~W. Thomas,
\newblock Phys.Lett. {\bf B215}, 141 (1988).

\bibitem{Bernard:2007zu}
V.~Bernard,
\newblock Prog.Part.Nucl.Phys. {\bf 60}, 82 (2008).

\bibitem{Bass:2009ed}
S.~D. Bass and A.~W. Thomas,
\newblock Phys.Lett. {\bf B684}, 216 (2010).

\bibitem{Bijker:2009up}
R.~Bijker, E.~Santopinto and E.~Santopinto,
\newblock Phys.Rev. {\bf C80}, 065210 (2009).

\bibitem{Bernard:2001rs}
V.~Bernard, L.~Elouadrhiri and U.-G. Meissner,
\newblock J. Phys. G {\bf G28}, R1 (2002).

\bibitem{Jain:1993qh}
P.~Jain and H.~J. Munczek,
\newblock Phys. Rev. {\bf D48}, 5403 (1993).

\bibitem{Maris:1999nt}
P.~Maris and P.~C. Tandy,
\newblock Phys. Rev. {\bf C60}, 055214 (1999).

\bibitem{Ball:1980ay}
J.~S. Ball and T.-W. Chiu,
\newblock Phys. Rev. {\bf D22}, 2542 (1980).

\end{thebibliography}

\end{document}